# Search for Magnetic Rotation in the A≈140 region

TERESA RZĄCA-URBAN

Institute of Experimental Physics
Warsaw University, PL-00-681 Warsaw, Poland

Magnetic dipole bands have been identified in $^{142,143,144}$Gd and $^{141}$Eu nuclei. The observed bands are based on configurations with small oblate deformation involving high-$\Omega$, $h_{11/2}$ proton particles and low-$\Omega$, $h_{11/2}$ neutron holes. Information about the strength of the effective interaction between the constituent nucleons has been deduced using a semiclassical scheme of the coupling of two angular momentum vectors. The experimental results are compared with predictions of tilted axis cranking calculations. The transition from irregular dipole band in $^{144}$Gd to more regular ones in lighter Gd nuclei is observed.

## 1. Introduction

Magnetic Rotation (MR) has been discovered in the neutron-deficient Pb nuclei with mass number around 200 [1]. This new type of rotation, observed in nearly spherical nuclei, is represented by rotation of a large magnetic dipole around the nuclear spin. At the bandhead, corresponding to such structure, angular momentum vectors of a few high-j valence protons in a stretched coupling and a few of high-j neutron holes in a stretched coupling (or vice-versa) are oriented perpendicularly to each other. The angular momentum of the band is generated by the gradual alignment of both spins along the total angular momentum vector. This process is called *shears mechanism* by analogy with the closing of a pair of shears, as illustrated in Fig.1. Magnetic Rotation is expected in weakly deformed nuclei in various places of the nuclear chart [2], where one has high-j particles of one kind of nucleons and high-j holes of the other kind. The active orbitals involved must have high j-values to give rise to long bands. To date bands corresponding to MR have been found in different nuclei in four mass regions (see [3] and references therein). Our studies concentrated on nuclei with proton number around $Z = 64$ and neutron number just below $N = 82$.

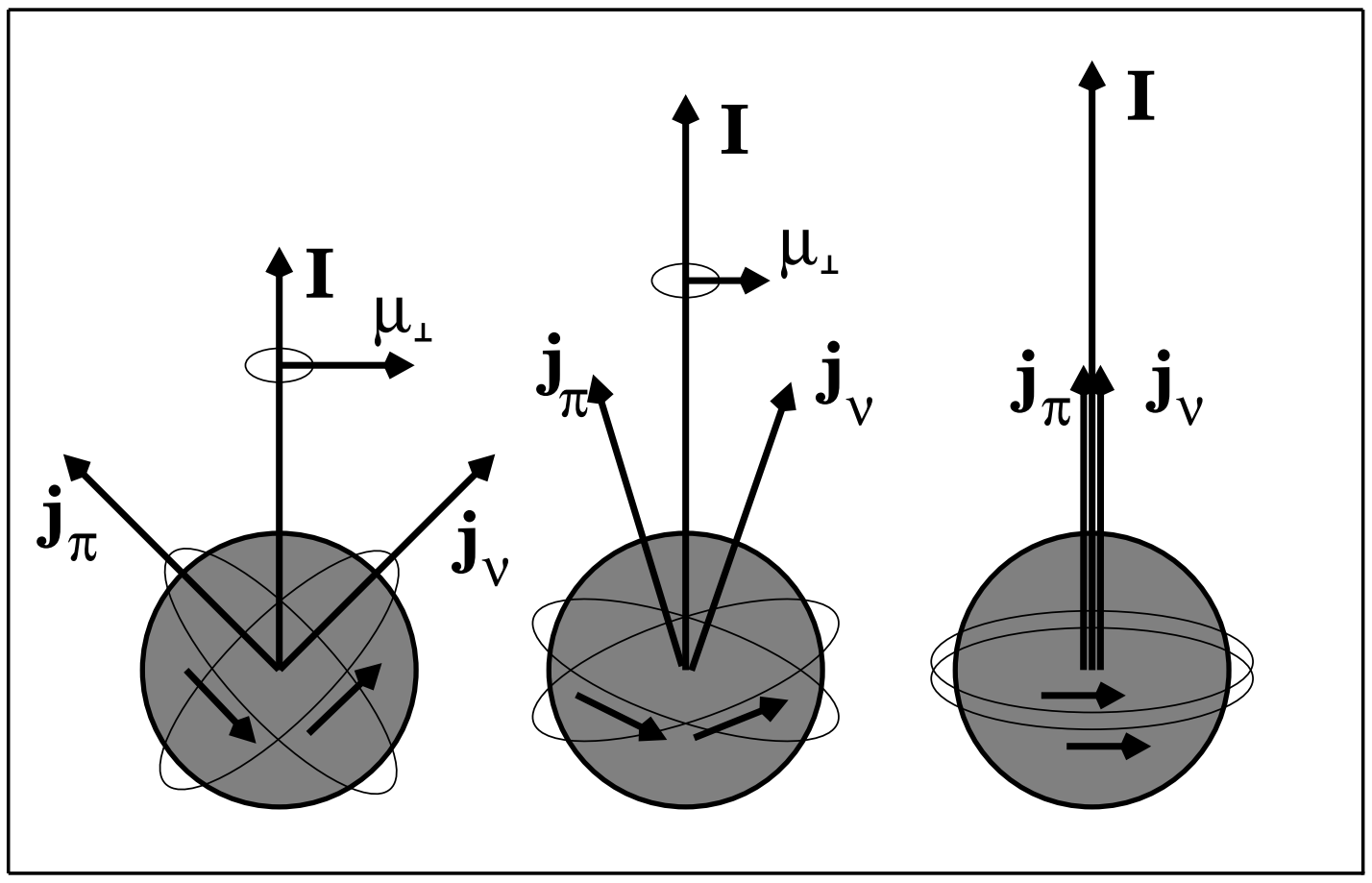

Fig. 1. Schematic drawing of the angular momentum coupling of neutron-hole and proton-particle blades in a shears band. From the left to the right, the three cases drawn, correspond to the beginning, the middle and the end of the band.

Low-lying excited states in these nuclei show irregular structures typical for nuclei with small deformation. MR bands in these nuclei would correspond to configurations involving high-$\Omega$, $h_{11/2}$ protons and low-$\Omega$, $h_{11/2}$ neutron holes. Magnetic Rotation has been already reported in $^{139}$Sm nucleus from this mass region [4].

## 2. Experimental details

Our search for superdeformed (SD) bands in Gd isotopes with A$\leq$146 provided extensive high-fold $\gamma$ coincidence data for several nuclei in this mass region. The data were reanalyzed in order to look for the predicted magnetic dipole bands in these nuclei. The results presented below have been obtaind using three different fusion evaporation reactions. High-spin states in $^{144}$Gd were populated in the $^{108}$Pd($^{40}$Ar, 4n) reaction at a beam energy of 182 MeV. The beam was delivered by the VICKSI accelerator of the Hahn-Meitner-Institue, Berlin. The $\gamma$-rays were measured with the OSIRIS spectrometer, consisting of 12 Compton-suppressed Ge detectors and an inner ball of 48 BGO scintillators, acting as a $\gamma$-ray multiplicity and sum-energy filter. Experiments with a thin- and a backed-target were carried out. Details of the experiment and data analysis were described in previous work [5]. The nucleus $^{143}$Gd has been investigated with the $\gamma$-spectrometer GASP and the charged-particle detector array ISIS at the Laboratori Nazionali di Legnaro, Italy. High-spin states in $^{143}$Gd were pop-ulated in the $^{97}$Mo($^{51}$V,p4n) reaction at a beam energy of 238 MeV. More details about the experiment were described in a paper reporting the identification of the SD band in this nucleus [6]. A study of $^{141}$Eu and $^{142}$Gd has been carried out with the $\gamma$-detector array EUROBALL III and the charged-particle detector array ISIS at the LNL, Italy. To populated high spin states in these nuclei the $^{99}$Ru($^{48}$Ti,xpyn) reaction at a beam energy of 240 MeV has been used. The $^{99}$Ru target, enriched to 95%, consisted of four self-supporting metal foils with a total thickness of 0.84 mg/cm$^2$. Experimental and data analysis details will be published later.

In all experiments the multipolarities of the $\gamma$-transitions were investigated by determinig the directional correlations (DCO ratios) for o $\gamma$-rays emitted from the de-exciting, orinented states. In addition, the linear polarization of strong $\gamma$-transitions was measured using CLOVER detectors of the EUROBALL III, working as polarimeters.

## 3. Results

Four dipole bands have been observed in $^{142}$Gd which are displayed in the partial level scheme shown in Fig.2. Three of them (1 to 3) were seen in the previous study of this nucleus [7] but, except for band 3, they were not well established. ¿From the present, high-statistics results detail information about these dipole bands has been obtained. Our spin-parity assignments are based on the angular distributions, DCO ratios, linear polarizations of $\gamma$-rays, conversion coefficients deduced from coincidence intensities and the observed decay patterns. For band 1 the order of the transitions as well as the deexcitation pattern are different than proposed in Ref. [7]. Three new, weak crossover transitions have been found, which determine the order of dipole transitons. The $\Delta$I=1 transitions have the E2/M1 mixing ratio $\delta$ =– 0.1±0.1. Ratios of the reduced transition probabilities B(M1)/B(E2) are large ($\geq$60 $\mu_N^2/e^2b^2$) for this band. Our band 2 consists of six $\Delta$I=1 transitions whereas in Ref. [7] only two have been proposed. Band 3 agrees with the previously published band. One dipole transition has been added at the lower end of the cascade and two additional crossover transitions have been found. Angular distributions and linear polarizations of the transition in this cascade indicate their $\Delta$I=1 and give an E2/M1 mixing ratio of $\delta$ =–0.1 ±0.1. Values of B(M1)/B(E2) for bands 2 and 3 are larger than 10 $\mu_N^2/e^2b^2$. The new band (band 4) consists of five in-band and four crossover transitions. It deexcites in a complex way into band 3.

The previously known level scheme of $^{141}$Eu [8] is revised and substantialy extended in the present work. Three dipole cascades have been found in our data. They are shown in Fig.3. A complete level scheme will be reported separately. Two of the previously known bands (band 1 and band 2) have been extended to higher energies and crossover E2 transitions were observed, confirming the order of $\gamma$-rays in a cascades. The DCO ratios obtained for $\gamma$-transitions in these cascades are consistent with a stretched-dipole character for all transitions in bands. At a higher excitation energy we have found a new band (band 3) starting at spin 39/2, which consists of seven in-band and two crossover transitions. The proposed spin assignments for this band are based on DCO ratios obtained for the transitions linking band 3 with low-spin states. Their uncertainities are large and the spin assignments are therefore tentative. For all bands the B(M1)/B(E2) ratios have been deduced wherever the crossover transitions were observed. For cases where the E2 transitions were not observed the lower limits for the B(M1)/B(E2) ratios were estimated, giving B(M1)/B(E2) ratios large than $10(\mu_N/\mathrm{eb})^2$.

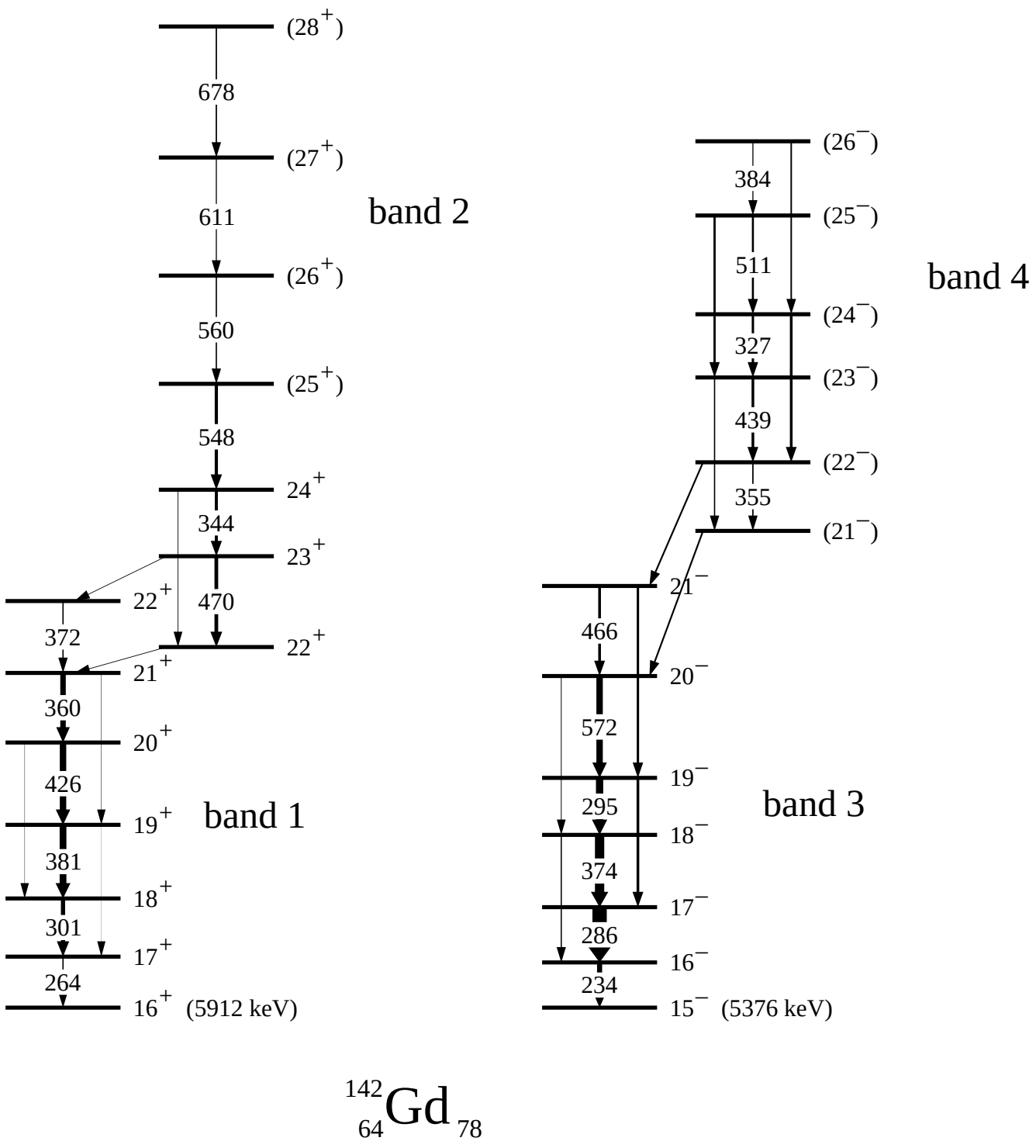

Fig. 2. Partial level scheme of $^{142}$Gd, showing four dipole bands observed in this nucleus.

The level scheme of $^{143}$Gd has been considerably extended in excitation energy and spin with respect to previous publication [9]. Partial level scheme of $^{143}$Gd is displayed in Fig.3 showing sequences which are of a relevance for the present paper. A few transitions from band 2 have been known before [7] but only our high-statistics data allowed to locate them in the level scheme. In addition, few new transitions belonging to this band have been found. Band 3 has been extended by four new dipole transitions. Most of the E2 cross-over transitions are not observed in these bands. The experimental lower limits for ratios of the reduced transition probabilities, B(M1)/B(E2), are $\geq 10(\mu_N/\mathrm{eb})^2$.

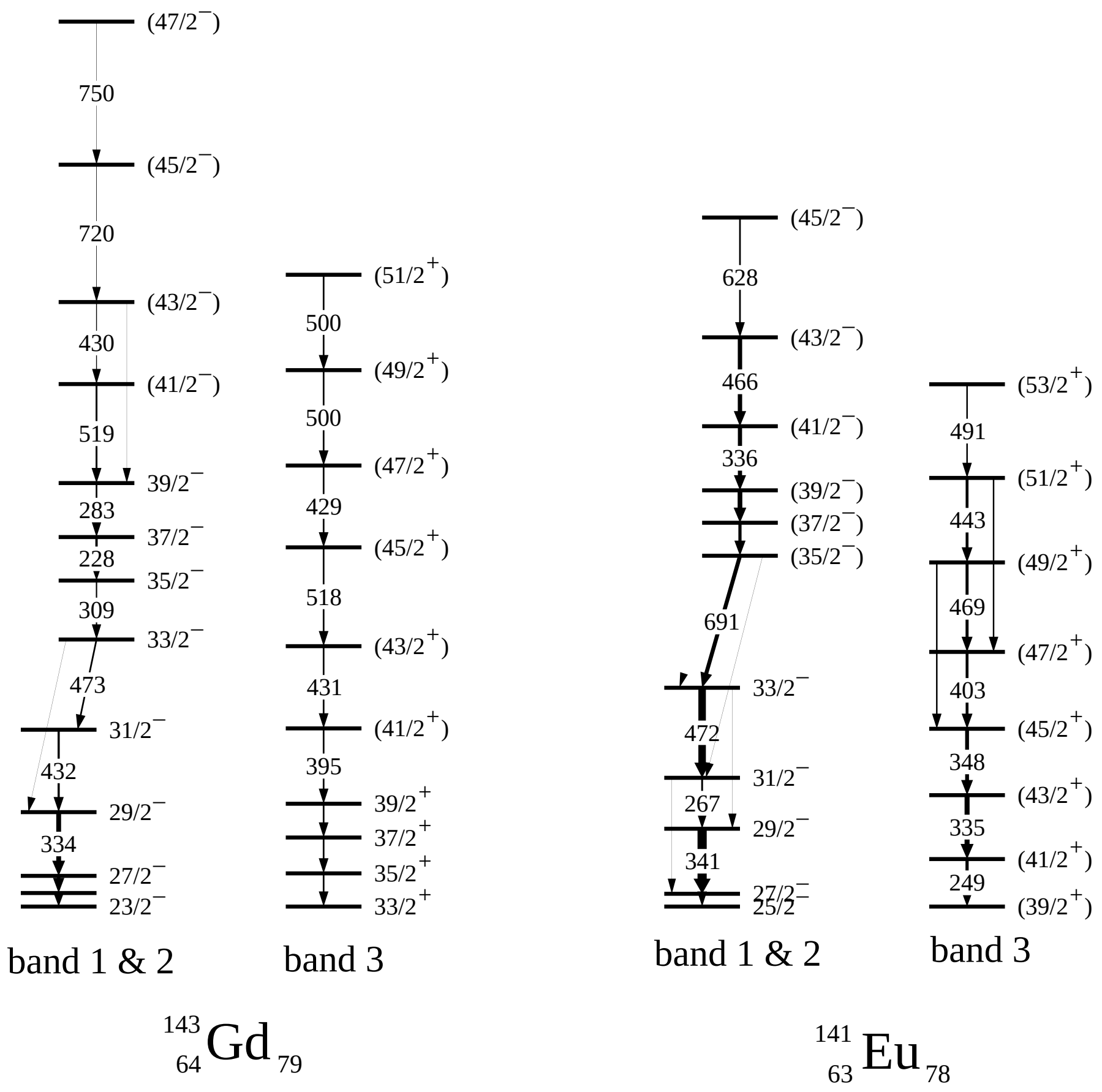

Fig. 3. Partial level scheme of $^{143}$Gd and $^{141}$Eu, showing dipole bands observed in these nuclei.

A dipole band in $^{144}$Gd has been reported in our previous paper [5].

For this band the angular momentum, drawn as a function of rotational frequency, has been compared in Fig.4 with similar plots for bands in other nuclei. All bands observed in $^{142,143,144}$Gd and $^{141}$Eu nuclei have similar features to MR bands identifed in other mass regions. The bands consist of strong magnetic dipole transitions with weak or unobserved E2 crossover transitions. In most of the bands (the $^{144}$Gd case will be discussed later) the in-band states follow a rotational-like pattern ( i.e. a near linear dependence of angular momentum vs. rotational frequency), as can be seen in Figs.4 and 5. The measured B(M1)/B(E2) ratios of reduced transition probabilities are large. These similar features suggest that the dipole bands in this mass region may also correspond to the Magnetic Rotation.

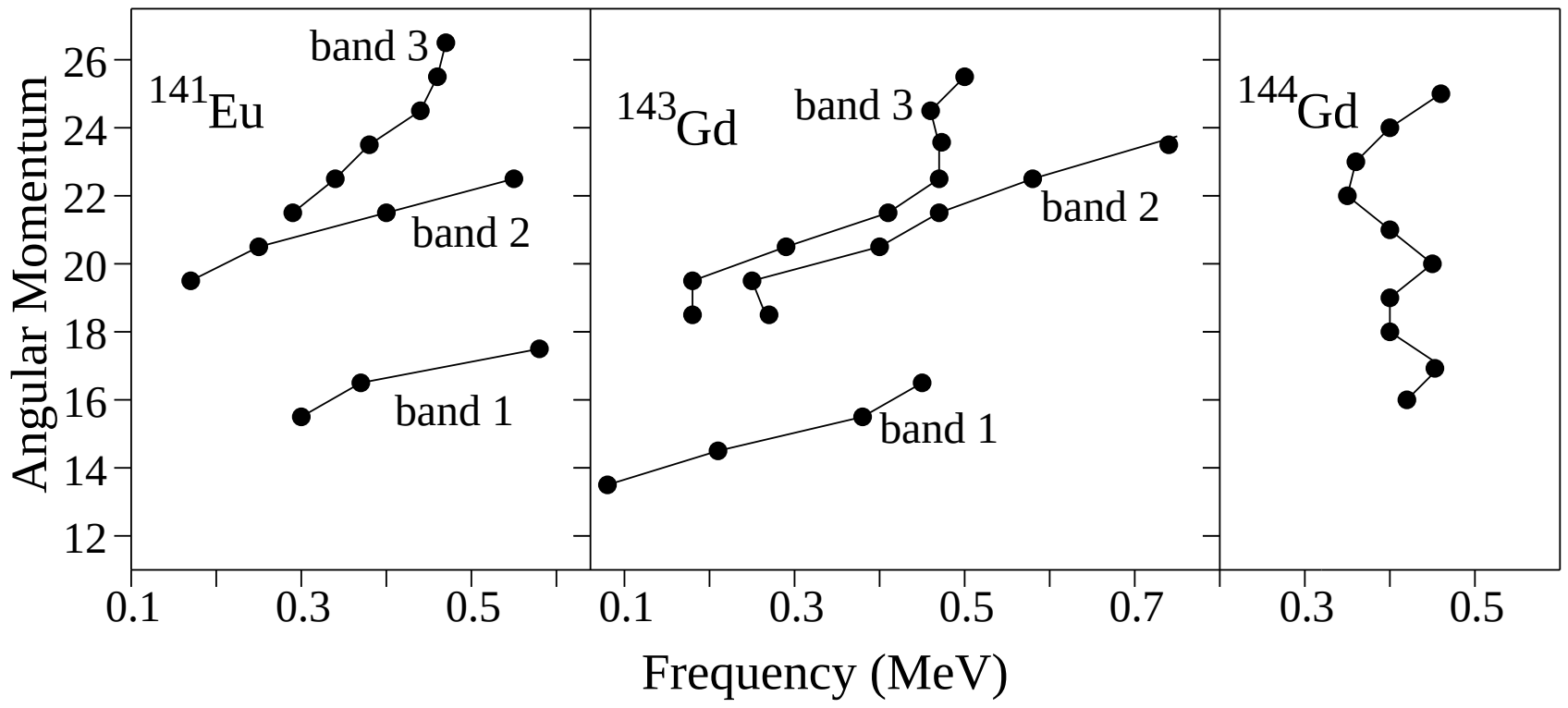

Fig. 4. Angular momentum as a function of rotational frequency for dipole bands in $^{141}$Eu and $^{143,144}$Gd.

## 4. Discussion

To investigate which configurations play a role in the observed M1 bands Total Ruthian Surfaces have been calculated using the cranked shell model. The calculations show minima for rather small oblate deformation in frequency range from 200 keV to 500 keV [10]. The experimental mixing ratios, $\delta$, found for the strongest $\Delta$I=1 transititons have small negative values implying a negative sign for the quadrupole moment and confirming oblate deformation. For this deformation the $h_{11/2}$ neutron-hole and $g_{7/2}$ proton-hole states with low-$\Omega$ value and the $h_{11/2}$ proton states with high-$\Omega$ value can contribute to the configurations of the observed dipole bands. In $^{142}$Gd the configuration assignment for the observed bands has been made based on comparison of experimental data with tilted axis cranking model (TAC) calculations [11, 12]. The $\nu h^{-2}_{11/2} \pi h^{2}_{11/2}$ configuration may be assigned to band 1 and $\nu h^{-2}_{11/2}\pi h^{1}_{11/2}g^{-1}_{7/2}$ configuration to band 3. Bands 2 and 4 likely result from the breakup of a second $h_{11/2}$ neutron-hole pair. The results of the calculations are shown together with experimental results in Fig.5 where the angular momentum as function of the rotational frequency is shown for these bands. The results of the calculations are indicated with dashed lines. They have similar slopes as the experimental curves, represented by solid lines, but lie a little bit above them. For two bands in the upper pannel calculations are not performed to higher frequency since it becomes difficult to follow the assumed configurations. The calculations were able to reproduce the experimental B(M1)/B(E2) ratios of the dipole bands resonably well, suggesting the applicability of the concept of the MR to these bands.

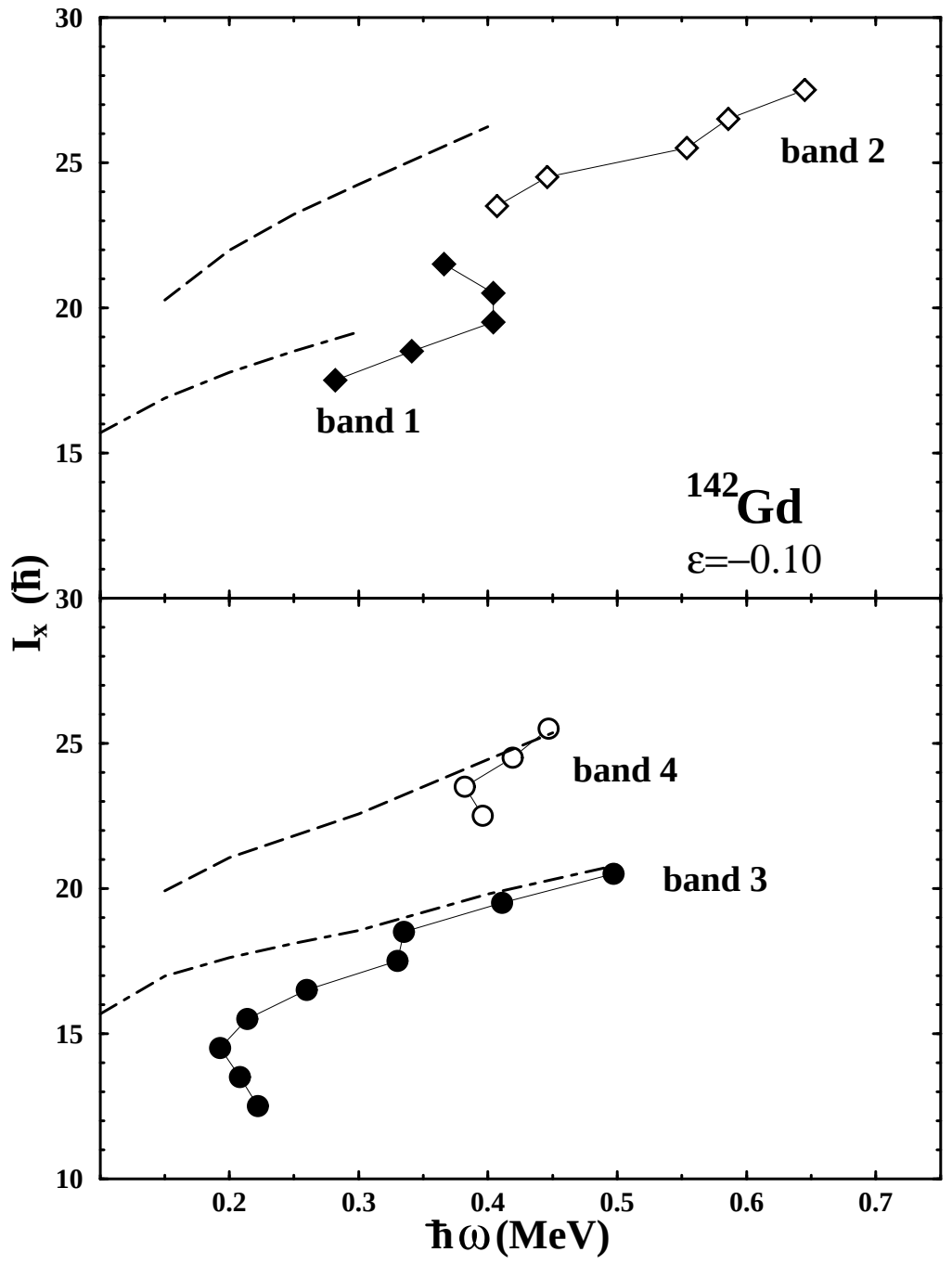

Fig. 5. Angular momentum as a function of rotational frequency for dipole bands in $^{142}$Gd.

The angular momenta of the dipole bands 1 and 2 in the $^{141}$Eu and $^{143}$Gd nuclei have similar frequency dependences as in $^{142}$Gd but are approximately $4\hbar$ smaller (see Fig. 4 and 5). This indicates that their configurations result from those of $^{142}$Gd by subtraction of a proton particle and a neutron hole, respectively. Hence, the $\nu h^{-2}_{11/2}\pi\ h^{1}_{11/2}$ configuration may be assigned to band 1 in $^{141}$Eu and $\nu h^{-1}_{11/2}\pi\ h^{2}_{11/2}$ configuration to band 1 in $^{143}$Gd. Bands 2 in both nuclei results from the breakup of a second $h_{11/2}$ neutron-hole pair. A configuration assignment to bands 3 in $^{141}$Eu and $^{143}$Gd, appears more difficult. Additional TAC model calculations are needed to determine the structure of these two bands.
The dipole band in $^{144}$Gd has been interpreted as $\nu h^{-2}_{11/2}\pi\ h^{n}_{11/2}(g_{7/2}/d_{5/2})^{-n}$ configurations in our previous paper [5].

In addition to the TAC analysis discussed above, observed bands 1 and 2 in $^{142,143}$Gd were analyzed with the simple, semiclassical approach of Macchiavelli et al. [13, 15, 14], which is based on a schematic model of the coupling of two long vectors, $\vec{j_\pi}$ and $\vec{j_\nu}$. The shears angle, $\theta$, between $\vec{j_\pi}$ and $\vec{j_\nu}$ for a given state with the total angular momentum $\vec{I}=\vec{j_\pi}+\vec{j_\nu}$, can be derived using the simple expresion:

$$\cos(\theta(I))=(I(I+1))-j_\nu(j_\nu+1)-j_\pi(j_\pi+1))/\ 2(j_\nu(j_\nu+1)j_\pi(j_\pi+1))^{1/2}$$

The semiclassical approach assume that the interaction between nucleons forming blades is independent on the shears angle $\theta$. In this way the energy needed to form each of the baldes is included in the bandhead energy and the excitation energy along the band is given only by the change in the potential energy, caused by the recoupling of the angular momenta in the shears. The effective interaction between the neutron-hole blade and the proton-particle blade is equal to the excitation energy calculated relative to the energy of the bandhead:

$$V(I(\theta)) = E(I) - E_{\mathbf{bandhead}}.$$

Figure 6 shows relative energies of states as a function of shears angle for band 2 in $^{142}$Gd. The effective interaction between the valence protons and neutrons can be expanded in terms of even Legendre polynomials. For bands observed in $^{142,143}$Gd significant part of the interaction between the shears blades may be represented by the $P_2$ term, as in Pb region. Relative energies of states as a function of the shears angle were fitted well (see Fig.6) with a function of the form:

$$V(\theta) = V_0 + V_2(3\cos^2\theta\text{-}1)/2$$

The strength of the effective interaction is determined here by the $V_2$ value. For bands 1 and 2 in $^{143}$Gd the maximum observed spin for each configuration is equal to the maximum that can be generated by closing the shears blades, so the core contribution is negligible. For bands in $^{142}$Gd a small effect due to the possible increasing contribution from the core is taken into account by decomposing the total spin as $I=I_{\mathbf{shears}}+R_{\mathbf{core}}$ and using a linear relation $R_{\mathbf{core}}=(\Delta R/\Delta I)(I-I_{\mathbf{band}})$, where $\Delta R$ is determined from the difference between the maximum observed spin and the sum of $j_\pi$ and $j_\nu$.

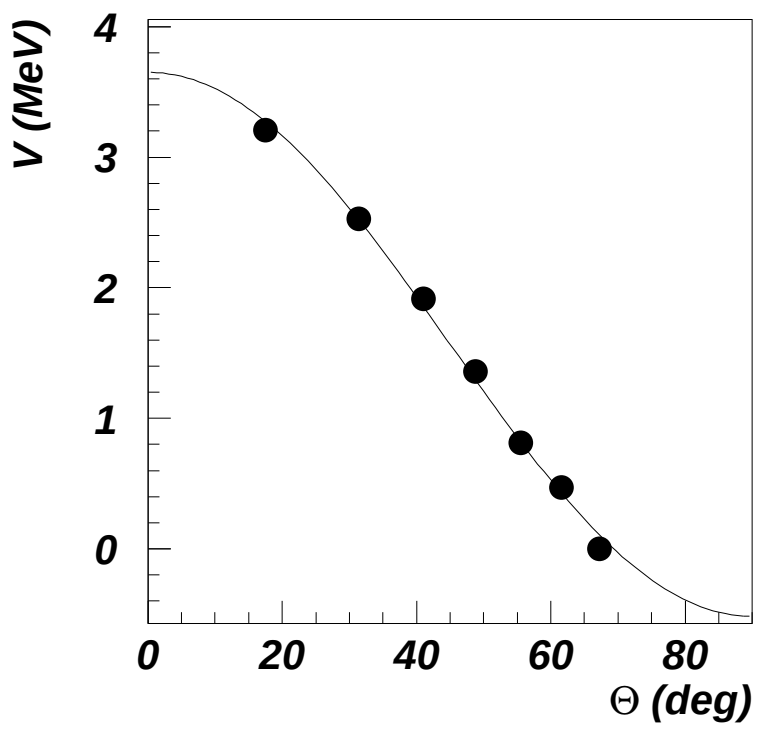

Fig. 6. The effective interaction between angular momentum vectors $\mathbf{j}_\pi$ and $\mathbf{j}_\nu$, as a function of the shears angle $\theta$. The solid curve is the expected dependence of a pure $P_2$ term in the interaction.

In our case this contribution is less than 10% of the total angular momentum at the top of the bands. The obtained strength of the interaction in band 2 in $^{142}$Gd is $V_2$=1724 keV. Taking into account two protons and two neutron holes as constituents of the blades we estimate that the interaction strength per proton/neutron hole pair is about 430 keV. Similar values of the interaction strength per proton/neutron hole pair are ∼350 keV, ∼400 keV and ∼ 380 keV for band 1 in 142Gd, band 1 in 143Gd and band 2 in 143Gd, respectively. These values should be compare to the value of ∼ 300 keV for $^{198}$Pb and ∼ 500 keV for $^{110}$Cd [14]. Our values are consistent with the 1/A dependence of the interaction strength suggested in [14].

The angular momenta of the observed bands as function of rotational frequency have been compared to each other in Figs.4 and 5 to find out if the properties of dipole bands depends on numbers of valence protons and neutrons. The dipole band in $^{144}$Gd shows a much less regular behaviour than other bands since the angular momentum increases within a narrow frequency range of about 0.1 MeV. The transition from the irregular dipole band in $^{144}$Gd to more regular ones in other nuclei results from an increase of the distance to the semi doubly-magic nucleus $^{146}$Gd. Similar transition has been observed in Cd, In and Sn nuclei with N approaching 50 (see [2] and references therein). This behaviour suggests that quadrupole polarizability is required to produce more regular dipole bands. The deformation induced by the valence particles and holes helps to align their angular momenta into two stable blades of the shears. On the other hands in the immediate vicinity of doubly-magic nuclei the short range residual interactions between valence particles dominate, which do not favour the formation of stable shears blades.

## 5. Conlusions

In conclusion, we have observed dipole bands in $^{142-144}$Gd and $^{141}$Eu nuclei with properties similar to those seen in the Pb region. They consist of strong M1 transitions with weak or unobserved E2 crossover transitions and in most of them the states follow a rotation-like pattern. TAC calculations suggest the bands are based on configurations with small oblate deformation involving high-$\Omega$, $h_{11/2}$ protons and low-$\Omega$, $h_{11/2}$ neutron holes. Interpretations of these bands are hampered by the absence of lifetimes for the levels of these bands. Therefore, more experimental data are needed to determine the B(M1) and B(E2) values.

## 6. Acknowledgments

The results presented in this paper come from a collaboration involving groups from IKP Jülich,LNL Legnaro, Universities of Warsaw, Bonn and Padova. The work was partly funded by the KBN Grant No. 5P03B 08420.